# Inside the Black Box: Detecting and Mitigating Algorithmic Bias Across Racialized Groups in College Student-Success Prediction

**Denisa Gándara**

*The University of Texas at Austin*

**Hadis Anahideh**

*University of Illinois, Chicago*

**Matthew P. Ison**

*Northern Illinois University*

**Lorenzo Picchiarini**

*University of Illinois, Chicago*

*Colleges and universities are increasingly turning to algorithms that predict college-student success to inform various decisions, including those related to admissions, budgeting, and student-success interventions. Because predictive algorithms rely on historical data, they capture societal injustices, including racism. In this study, we examine how the accuracy of college student success predictions differs between racialized groups, signaling algorithmic bias. We also evaluate the utility of leading bias-mitigating techniques in addressing this bias. Using nationally representative data from the Education Longitudinal Study of 2002 and various machine learning modeling approaches, we demonstrate how models incorporating commonly used features to predict college-student success are less accurate when predicting success for racially minoritized students. Common approaches to mitigating algorithmic bias are generally ineffective at eliminating disparities in prediction outcomes and accuracy between racialized groups.*

Since the emergence of "big data" in the 1990s, efforts to use advanced statistical techniques to predict outcomes of interest have proliferated in various social domains, education notwithstanding (Baker et al., 2019; Government Accountability Office [GAO], 2022). The suite of techniques used to forecast outcomes and inform decision-making within organizations is broadly known as "predictive analytics." Although largely unseen, predictive analytics fuel myriad decisions within educational institutions, from college admissions (Hutt et al., 2019) and student retention interventions (Baker et al., 2019) to resource allocation (Wayt, 2019; Yanosky & Arroway, 2015). Evidencing the pervasiveness of predictive analytics, in a survey of nearly 1,000 colleges and universities, 89% of respondents reported making some investment in predictive analytics (Parnell et al., 2018).

A key component within the vast array of predictive statistical techniques is the predictive model, a computational tool that maps the input set of attributes of individuals (e.g., high school GPA and demographic features) to their outcomes (e.g., college credits accumulated) in order to identify underlying associations and patterns in the data. The predictive model is especially useful with large datasets, where it is impossible or inefficient to identify associations and patterns manually.

In recent years, observers have raised concerns that predictive models in education may perpetuate social disparities (GAO, 2022). For instance, a model that includes socially relevant attributes, such as race, gender, or income, will often predict that students from socially disadvantaged categories (e.g., women in STEM) will have less favorable outcomes. Such a model extrapolates from prior relationships between socially relevant attributes (e.g., race) and educational outcomes (e.g., graduation) that are partly shaped by societal injustices, such as racism, sexism, and classism (e.g., López et al., 2018).

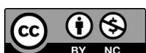

In this study, we appraise predictive models within the higher education context by examining: (1) how predictions of college student success differ between racial/ethnic groups, (2) how model performance (accuracy) differs between racial/ethnic groups, and (3) the effectiveness of common techniques to mitigate algorithmic bias. While researchers and observers have noted that predicted outcomes may differ between racial groups (Barocas & Selbst, 2016; Ekowo & Palmer, 2017), it is also important to understand how model *performance* varies between groups. That is, do models predict success more accurately for some racial/ethnic groups than others? This could occur if, for instance, the data for racially minoritized students contain more errors, if models are missing variables that are more predictive of success for racially minoritized students, or if racially minoritized students with successful outcomes are underrepresented in the historical data used to train the models. Moreover, it is important to examine the extent to which common techniques for mitigating algorithmic bias are effective at reducing disparities in prediction outcomes and accuracy between racial/ethnic groups.

The reason for focusing on racialized disparities is that educational attainment rates across racial/ethnic groups remain markedly unequal (U.S. Department of Education, 2021). Given these inequities in educational attainment levels, predictive models that are agnostic to racial bias may penalize groups that have been subject to racialized social disadvantages.

We situate our statistical analyses within relevant historical and social contexts (Zuberi, 2001), recognizing that racially minoritized groups are disadvantaged in the educational context through various interlocking social systems of oppression (Collins, 2000; Reskin, 2012). Although an exhaustive review is beyond the scope of this paper, we refer readers to examples of systems, structures, and practices that penalize racially minoritized groups. In the education domain, oppressive barriers to educational success include educational tracking (Oakes, 1985), deepening school segregation (Orfield et al., 2012), teacher racial bias (Gershenson & Papageorge, 2018), racial disparities in school funding that track with levels of segregation (Weathers & Sosina, 2022), and disparate punishment of Black and Latinx students (Davison et al., 2022). Racially minoritized students' educational success is also conditioned by racialized barriers outside education, including constraints on wealth accumulation and income, which limit students' ability to pay for higher education (Mitchell et al., 2019).

It is important to understand this background since the state of the world, which is rooted in various societal injustices, affects the data distribution. These historical injustices condition educational opportunities and experiences for racially minoritized students. Subsequently, when predictive models make predictions on students who are racially minoritized, they may be predicted to fail, reinforcing historical injustices. In this context, this study addresses the following questions:

1. To what extent do college student success predictions differ between racial/ethnic groups?
2. To what extent does the accuracy of college student success predictions differ between racial/ethnic groups?
3. How effective are computational strategies in mitigating bias in predictions across racial/ethnic groups?

While "bias" can be conceptualized in myriad ways, we define "bias" using four widely used statistical measures of algorithmic (un)fairness (Barocas et al., 2017; Pessach & Schmueli, 2022). The first of these notions measures disparities in prediction outcomes between racial/ethnic groups, while the other three measure disparities in model performance (or prediction *accuracy*) between racial/ethnic groups. These measures are defined in greater detail in the next section.

Findings from this study show that models incorporating commonly used features to predict college student success predict worse outcomes for Black and Hispanic students. More importantly, these prediction models are less accurate when predicting college success for these racially minoritized groups. For example, they are more likely to predict failure for students who actually succeed if those students are categorized as Black or Hispanic. With respect to bias-mitigation techniques, those that modify the algorithms to incorporate fairness during the training process reduce disparities in predictions between racially advantaged and disadvantaged groups. However, no technique effectively eliminates disparities in prediction outcomes or accuracy across any notion of fairness.

**Conceptualizing Fairness**

Before reviewing related literature on college student success prediction, we introduce relevant conceptualizations of algorithmic fairness or algorithmic bias. These technical terms, which emerge from the subfield of fairness-aware machine learning, refer to specific statistical notions (defined in a subsequent subsection). We recognize a distinction between statistical notions of algorithmic bias or (unfairness) and *practical* unfairness, whereby human uses of prediction algorithms may be considered "unfair." For instance, a model that predicts worse outcomes for racially minoritized students would be considered "unfair" according to one statistical definition of algorithmic fairness but is not necessarily unethical or unjust. Such a model could, for example, lead to financial aid distributions that benefit racially minoritized students more, which may be justified given these students' generally greater sensitivity to price (Heller, 1997). Unless otherwise noted, in this paper, the terms "bias," "fairness," and "unfairness" refer to statistical measures associated with prediction algorithms rather than practical or philosophical notions of fairness.

### *Individual Versus Group Fairness*

Individual fairness and group fairness are two different approaches to measuring fairness in algorithmic decision-making. Individual fairness focuses on treating individuals fairly on a case-by-case basis. In other words, it aims to ensure that similar individuals are treated similarly, regardless of their membership in any particular group. For example, if two individuals have similar academic records and are both at risk of dropping out, an individual fairness approach would ensure that both receive similar levels of support, regardless of their gender, race, or any other characteristic. Group fairness, on the other hand, focuses on ensuring that groups of individuals are treated fairly as a whole. It aims to address systemic, rather than interpersonal, biases and inequities that may exist within different demographic groups.

In this work, we are interested in systemic discrimination and oppression that leads to differential access to educational opportunity—and other resources (e.g., wealth and homeownership) that (re)produce educational advantages—across racial/ethnic groups. Since our focus is on identifying and mitigating *systemic* racial disparities, we employ group fairness metrics. Group fairness metrics are designed to assess and quantify fairness at the group level, such as ensuring that the outcomes of an algorithm are consistent across different racial or ethnic groups. This can help identify and correct systemic biases that may be present in the algorithm or captured in the data, reflecting societal inequities. Moreover, by focusing on group fairness, we recognize that any real-world datasets and systems capture societal injustices that affect entire groups of people, such as racial or gender discrimination. In such cases, individual fairness may not be sufficient to address the root cause of the unfairness, whereas group fairness can help to identify and correct these systemic biases. The final reason for using group fairness measures is based on our interest in examining mitigation techniques since most of these techniques are based on group fairness notions.

### *Fairness Measures*

Specifically, we employ four widely accepted notions of group fairness to evaluate the fairness of student success predictions across racial/ethnic groups: statistical parity, equal opportunity, predictive equality, and equalized odds (Barocas et al., 2017; Pessach & Schmueli, 2022). Although scholars have considered other notions of fairness, we focus on these four because current bias-mitigation algorithms, which we assess in this study, are designed based on these core notions and other criteria derived from them (Pessach & Schmueli, 2022). Therefore, these four fairness criteria provide a practical approach for evaluating the performance of bias-mitigation algorithms in real-world applications. In practice, users can select the measure of fairness that is preferred based on context, knowledge of social disparities, use case, and regulations. We briefly describe each algorithmic fairness notion in turn; the probabilistic definitions of these notions appear in the supplemental materials (Appendix A). It should be noted that the original definitions were established using binary sensitive attributes (privileged versus unprivileged), which makes the absolute value of group probability differences a reasonable measure of the magnitude of unfairness. However, when examining fairness across multiple subgroups instead of just at the binary level, we retain the sign of the differences to determine which group had a lower probability than the other.

First, statistical parity is achieved by having equal favorable outcomes (degree attainment) received by the unprivileged group (e.g., Black) and the privileged group (e.g., White). Said differently, under the notion of statistical parity, we consider a model fair if being a member of a racially minoritized group is not correlated with the probability of bachelor's degree attainment.

The next three fairness measures build on the statistical notions of true/false positives/negatives (for a visual, see Confusion Matrix in Appendix B). Specifically,

- A true positive result would correctly predict success for a student who succeeds (in our case, attains a bachelor's degree).
- A true negative result would correctly predict failure for a student who does not succeed.
- A false positive result (Type I error) would incorrectly predict success for a student who does not succeed.
- A false negative result (Type II error) would incorrectly predict failure for a student who does succeed.

Building on these statistical notions, equal opportunity represents equal false negative rates between groups. This fairness notion requires that each group receives the negative outcome at equal rates, conditional on their success. In other words, under this notion, the model should (incorrectly) predict failure for those who succeeded (attained at least a bachelor's degree) at the same rate for students from different racial/ethnic groups. This notion assumes knowledge of the true outcome values (whether a student attained at least a bachelor's degree) and aims to satisfy parity between socially relevant groups, subject to the true values.

A third fairness notion is predictive equality, which represents equal false positive rates. To satisfy this criterion,

positive predictions (that a given student will attain a bachelor's degree) for students who do not actually attain a bachelor's degree should be the same across racial/ethnic groups.

Finally, the notion of equalized odds represents the average difference in false positive and true positive rates between groups. To achieve fairness under this notion, both the false positive rate (wrongly predicting success) and the true positive rate (correctly predicting success) should be the same across racial/ethnic groups. We measure fairness in college student success predictions using these four statistical notions.

**Related Literature on Fairness in College Student Success Prediction**

In recent years, educational researchers and data scientists have begun to develop insights into algorithmic (un) fairness and bias within various stages of the machine learning (ML) process. Generally, the goal of prediction models is to maximize the accuracy of the predictions, but research on algorithmic fairness examines not only the accuracy but also the fairness (or unfairness) of the prediction and the potential tradeoff between the two (Wang et al., 2021). Among the most important discernments from these studies are: (1) the importance of the representation of socially relevant groups in training datasets (Riazy et al., 2020),[1] and (2) novel statistical techniques intended to measure predictive fairness between groups (Gardner et al., 2019; Hutt et al., 2019).

A small but growing number of studies have examined algorithmic fairness in the domain of college student success (Anderson et al., 2019; Hu & Rangwala, 2020; Hutt et al., 2019; Lee & Kizilcec, 2020; Yu, Lee, & Kizilcec, 2021; Yu, Li, Fischer, 2020). Regardless of the data source or set of variables included, most of these studies have found higher false negative rates (predicted failure for students who actually succeeded) for racially minoritized students (Anderson et al., 2019; Lee & Kizilcec, 2020; Yu, Lee, & Kizilcec, 2021; Yu, Li, Fischer, 2020). Findings for other notions of fairness are more mixed (Lee & Kizilcec, 2020; Yu et al., 2020).

Anderson et al. (2019), who used administrative data from a single institution, found higher false positive rates and lower false negative rates for White students and the reverse for Hispanic/Latinx students. Similarly, Yu et al. (2020) examined how predictions of college outcomes differed according to the data source used (e.g., learning management system [LMS], institutional data, or survey data). They found higher false negative rates for at least one group of racially minoritized students (Black, Hispanic, or international students) across all data sources except click data (number of clicks in LMS; Yu et al., 2020). Likewise, using data from a large U.S. research university, Yu et al. (2021) detected disparities in predictions between "underrepresented minority" (URM) students and the aggregate group of Asian and White students on three metrics: prediction accuracy; recall, defined as "the proportion of actual dropouts who are correctly identified"; and the true negative rate, defined as "how likely a student who persists . . . is predicted to persist" (p. 5).

Lee and Kizilcec (2020) also detected disparities in student success predictions according to certain definitions of algorithmic fairness. In that paper, the authors evaluated predictive models for identifying at-risk students using three measures of statistical fairness: demographic parity, equality of opportunity, and positive predictive parity. They found that the models exhibit gender and racial bias in two of the three fairness measures considered.

Beyond exploring unfairness in ML models predicting college student success, our study tests various approaches for mitigating bias, both in data preparation (preprocessing) and in the models (in-processing). Previous work on bias mitigation within the domain of college student success prediction has focused on the impact of including or excluding sensitive attributes (e.g., race/ethnicity) in prediction models. For example, Yu et al. (2021) found that racialized disparities in college dropout predictions largely persisted regardless of whether the models contained sensitive attributes, including whether the student was categorized as URM. Specifically, the accuracy of predictions was unaltered with the inclusion of sensitive attributes, and fairness was only modestly improved.

Similarly, while not the focus of their study, Bird et al. (2021) also found that including sensitive attributes in models did not influence model performance. That study evaluated how accuracy was affected when sensitive attributes were excluded from the model, finding negligible changes (less than 1%). Notably, these findings are consistent with those from Yu et al. (2021), which were set in a significantly different context (Virginia community colleges versus a large research university).

To our knowledge, only one prior study has evaluated the effectiveness of statistical techniques to mitigate racialized bias in college student success predictions (Hu & Rangwala, 2020). Specifically, Hu and Rangwala proposed a course-based prediction model to predict whether students who have not taken a series of courses prior to a target course will fail that course. Their approach differed considerably from ours, given their study's distinct aims. In particular, Hu and Rangwala used a metric-free approach for fairness consideration and fitted separate models for each sensitive group to calculate the prediction gaps. The individual fairness metric was then added as a penalty to the sum of squared errors to estimate the parameters of neural network models subject to fairness. The results show improved fairness and accuracy compared to baseline methods.

Hu and Rangwala's (2020) contribution differs from ours in three primary ways. First, that study used an individual fairness notion rather than examining group fairness.

As mentioned previously, given our interest in systemic injustices based on racialization, group fairness measures are most appropriate for our study. Second, Hu and Rangwala's paper employed a single ML model, neural networks, which is typically recommended for much larger datasets (Bengio, 2012) and is less transparent for decision-making purposes (Molnar, 2020). Third, Hu and Rangwala used a binary sensitive attribute, whereas we evaluated fairness across various racial/ethnic groups to examine how prediction outcomes and model performance differ between specific groups.

Our work extends previous studies in this area, particularly the important contribution of Hu and Rangwala, by evaluating techniques to mitigate fairness at the group rather than individual level. Moreover, we use a rich, nationally representative dataset. In addition to enhancing the generalizability of the findings, by using this public dataset, we allow for replication of this work and for comparability across studies to build a knowledge base about algorithmic fairness in college student success predictions. Furthermore, we bolster our empirical contribution by exploring various notions of fairness at the group level, presenting conceptual models that can be used for further exploration of unfairness in student success predictions.

## Data Sources

Data come from the Education Longitudinal Study of 2002 (ELS), a nationally representative, longitudinal study of students who were 10th graders in 2002. Given our focus on bachelor's degree attainment, the dataset is filtered based on the institution type to only include students who attended four-year postsecondary institutions. The outcome variable captures students' highest level of education as of the third follow-up interview (eight years after expected high school graduation). To construct a binary classification problem, we label students with a bachelor's degree and higher as the favorable outcome (label = 1) and all others as the unfavorable outcome (label = 0).

Predictive variables include features commonly used for student success prediction, including student demographic characteristics, socioeconomic traits, grades, and college preparation (Anderson et al., 2019; Attewell et al., 2022; Hutt et al., 2019; Yu et al., 2021).[2] We also include K12 school-level contextual variables that could be predictive of college student success (school enrollment, geographic region, percent of students on free or reduced-price lunch, and school control). Since category labels are not ordinal, we create binary variables for each level of the categorical variables following National Center for Education Statistics (NCES) (n.d.) documentation. The complete list of variables appears in the supplementary materials (Appendix C). Although our dataset does not include all possible variables that could be incorporated into a model that predicts college student success, our dataset has the advantage of being large ($n = 15,244$) and nationally representative and including the most commonly used features (27 predictors in total) based on our review of literature on college student success prediction.

Since we have a high number of missing values, we ran the models separately with multiple imputations (Rubin, 1996) and without imputations (listwise deleted rows with missing data).[3] To avoid the confounding impact of imputation on both unfairness and model performance, we stratified the response variable (bachelor's degree attainment) and racial groups for the training-testing splits, retaining the distribution of the historical data in both partitions. For simplicity, we present results without imputation in our main results. The results with imputation appear in the supplementary materials (Appendix D). A deeper investigation of how imputation affects the unfairness of the prediction outcome appears elsewhere (Nezami et al., 2024).

First, we randomly split the dataset into training and testing subsets with an 80:20 ratio (80% training, 20% testing). The ML models were trained on the training data and evaluated on the testing data to demonstrate their generalizability. To evaluate the prediction outcome using various fairness notions (described previously), we stratified the training and testing datasets by the outcome variable class labels (1, 0) and racial/ethnic categories, ensuring that we have enough observations from each group. The results were averaged over 30 different splits of the data. Table 1 presents the distribution of the outcome variable by racial/ethnic category after dropping observations with missing values.

## Analysis Methods

### Evaluating Unfairness

We employed four widely used ML models in higher education, including Decision Tree (Hamoud et al., 2018), Random Forest (Pelaez, 2018), Logistic Regression (Thompson et al., 2018), and SVM (Agaoglu, 2016). Each ML model has predefined parameters known as hyperparameters that must be provided before the training phase (e.g., depth of the tree in Decision Trees). Since the optimal values of such hyperparameters are data-dependent, we performed a five-fold cross-validation (CV) for each model to determine the best set of hyperparameters. In this process, the dataset was divided into five partitions, four of which were utilized for training and one for validation. Cross-validation repeats this process and selects a different partition for validation each time. A grid of feasible hyperparameters was assessed based on the CV schema described previously to choose the optimum set. Under 30 distinct random splits of training and testing datasets, we obtained the best set of hyperparameters before we performed model training. To evaluate model performance, we report the average and variance of the accuracy, as well as unfairness toward different racial/ethnic groups using various notions of unfairness.

TABLE 1

*Distribution of Bachelor's Degree or Higher Variable by Racial/Ethnic Category*

| Race | Bachelor's or Higher | % of data |
|---|---|---|
| Asian, Hawaiian/Pacific Islander | 1 | 0.72800 |
| | 0 | 0.27200 |
| Black or African American | 1 | 0.55851 |
| | 0 | 0.44149 |
| Hispanic | 1 | 0.63339 |
| | 0 | 0.36660 |
| More than one race | 1 | 0.63265 |
| | 0 | 0.36735 |
| White | 1 | 0.71768 |
| | 0 | 0.28232 |

### *Mitigating Bias*

In addition to evaluating unfairness in algorithms, we implemented statistical techniques to mitigate bias. Such techniques can be categorized into three groups: preprocessing, in-processing, and post-processing approaches (Pessach & Schmueli, 2022). Preprocessing techniques involve fairness evaluation in the data preparation step, which in turn should mitigate bias for downstream tasks. In contrast, in-processing techniques generally involve modifying the ML algorithms to account for fairness during the training process, such that the parameter estimation of the classifier forces the prediction outcome to be fair toward all (racial/ethnic) groups. The enforcement is accomplished in the optimization subproblem by adding a fairness metric as a constraint.

In this study, we employed two preprocessing and two in-processing mitigation techniques. We opted to exclude post-processing techniques for mitigating bias since these mechanisms are implemented at a later phase in the learning process, often producing inferior results (Woodworth et al., 2017). Post-processing bias-mitigation strategies also tend to be more controversial in practice, since they involve changing the prediction outcome after the model has been trained and are thus less likely to be used in education settings (Hirschman & Bosk, 2020).

From the array of bias-mitigation techniques, we selected reweighting, disparate impact remover, exponentiated gradient reduction, and meta-fair classifier based on their widespread application and potential to address distinct sources of bias in predictive modeling (Ferrara, 2023; Hort et al., 2023). By employing these techniques collectively, we sought to comprehensively address various sources of bias in our predictive modeling analysis.

We applied two preprocessing techniques: reweighting (Kamiran & Calders, 2012) and disparate impact remover (DIR) (Feldman et al., 2015). Reweighting assigns different weights to the training samples in each combination of racial/ethnic group and outcome-variable class label (e.g., Black X outcome label = 1). It does so before training a model to adjust the bias across groups. Because individual observations from the unprivileged groups with positive outcomes are underrepresented in the training data (see Table 1), classifiers are susceptible to bias. In this preprocessing approach, the data points representing successful outcomes for unprivileged groups are identified and upweighted to have a larger influence on model training.

In contrast to reweighting, DIR changes the distributions of other features in the model (not race/ethnicity) to force distributions to overlap at the group level. This process removes the ability to distinguish between group membership from a feature that otherwise offers a good indication to which group a data point may belong. As such, DIR directly targets disparate impact, where certain groups receive favorable outcomes more often than others, by adjusting feature values.

In addition to the two preprocessing techniques, we used two in-processing bias-mitigation strategies. The first in-processing technique is exponentiated gradient reduction (ExGR), which is designed to minimize the impact of sensitive features on the model's predictions (Agarwal et al., 2018). It does so by iteratively reweighting the training examples to decrease the model's sensitivity to sensitive attributes, such as race/ethnicity.

The second in-processing technique, meta fair classifier (Celis et al., 2018), takes a large class of fairness metrics as inputs and returns an optimal classifier that is fair with respect to constraints on the given set of metrics. This technique involves training a secondary model to adjust the predictions of the primary model based on sensitive attributes. It essentially acts as a corrective mechanism by learning to counteract the biases present in the primary model's predictions. By explicitly considering fairness metrics, metaclassifiers offer a fine-tuned adjustment to the base model's predictions, adapting to various types of bias. An advantage of this approach is that it works for various fairness criteria rather than a single fairness metric (e.g., Zafar et al., 2017).

In the results section, we refer to reweighting as ReW, disparate impact remover as DIR, exponentiated gradient reduction as ExGR, and metaclassifier as MetaC. For comparison, we also consider the baseline classification scenario, where no mitigation strategy is used.

### *Comparisons*

We used two comparison approaches to appraise model unfairness and test mitigation techniques—namely, (1) the subgroup level (i.e., each racial/ethnic group versus the rest) and (2) the aggregate level (i.e., privileged versus unprivileged). First, we compared each racial/ethnic group against all others and considered 1 for a certain group (e.g., Black) and 0 for every other group (e.g., White, Asian, Hispanic, and two or more races) to calculate gaps as discussed previously.

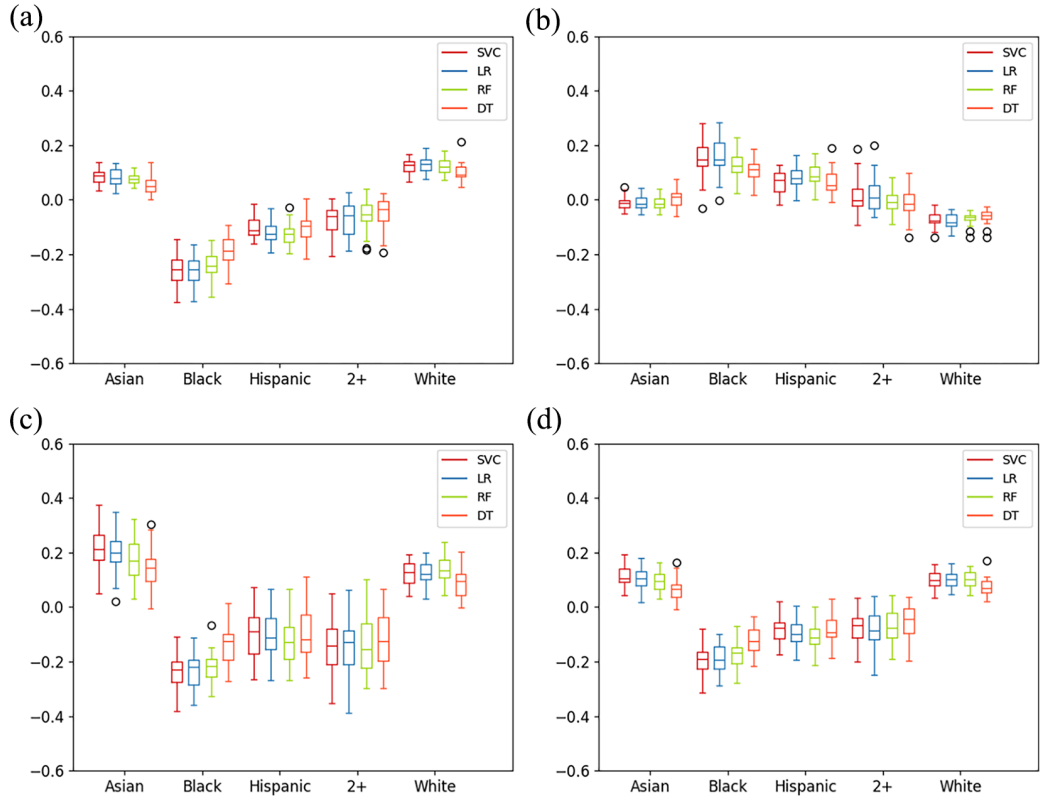

FIGURE 1. *Baseline with all ML models for all racial/ethnic groups. (a) Statistical parity of baseline for different ML models, (b) equal opportunity of baseline for different ML models, (c) predictive equality of baseline for different ML models, and (d) equalized odds of baseline for different ML models.*

To evaluate the limitations of data aggregation, which is common in this type of work, we also aggregated White and Asian groups in the privileged category and Black, Hispanic, and two or more race (2+) groups in the unprivileged category. While we recognize that Asian racialization is distinct and encompasses a diverse group of students, given the higher rates of educational attainment among this group in the aggregate, in this study, we consider this group statistically "advantaged" within educational contexts.

These comparisons represent an extension over prior work as they allow us to investigate the impact of existing mitigation techniques at both the subgroup and aggregate levels. Most existing techniques only work with binary sensitive attributes (e.g., "White" and "non-White"), requiring the researcher to specify the privileged group and forcing other subgroups to be aggregated as the unprivileged group (Pessach & Schmueli, 2022).

Although some existing unfairness mitigation techniques have the potential to incorporate nonbinary sensitive attributes, such extension has not been implemented in the literature. Binarizing sensitive attributes (1: privileged, 0: unprivileged) for the mitigation processes may not reduce fairness gaps for each group. This is important in educational settings, where research shows that students from different racial/ethnic groups have distinct experiences and outcomes (e.g., López et al., 2018). Therefore, it is critical to evaluate unfairness after applying mitigation techniques at the subgroup levels, as there may be significant differences between unprivileged subgroups.

## Results

We find no significant differences between the performance (accuracy) of different ML classifiers, although there are differences in unfairness levels between fairness notions and models (see Appendix E). To facilitate comparison, Figure 1 presents results for all ML models. We discuss the main findings for our assessments of unfairness and the effectiveness of bias-mitigation techniques in turn.

### Evaluating Unfairness

*Subgroup Level: Each Group Versus the Rest.* Figure 1 shows a comparison of unfairness levels using all four fairness notions and ML models for the baseline (without bias mitigation). The testing accuracy across these models is 77%, with a variance of 1.72% and an F1-score[4] of 75%, on average (Appendix E). These results indicate that models are generally biased against Black and Hispanic groups. Under the fairness notions of statistical parity (Figure 1a), predictive equality (Figure 1c), and equalized odds (Figure 1d), the boxes for Black and Hispanic students are at a lower level across all ML models, indicating that these students receive favorable outcomes (i.e., bachelor's attainment or higher) at a lower rate than students in other categories. For the notion of equal opportunity (Figure 1b), higher levels in the box plots, which we observe for Black and Hispanic groups, represent more unfairness. Specifically, these higher levels indicate that the prediction models are more likely to predict failure for Black and Hispanic students who succeeded.

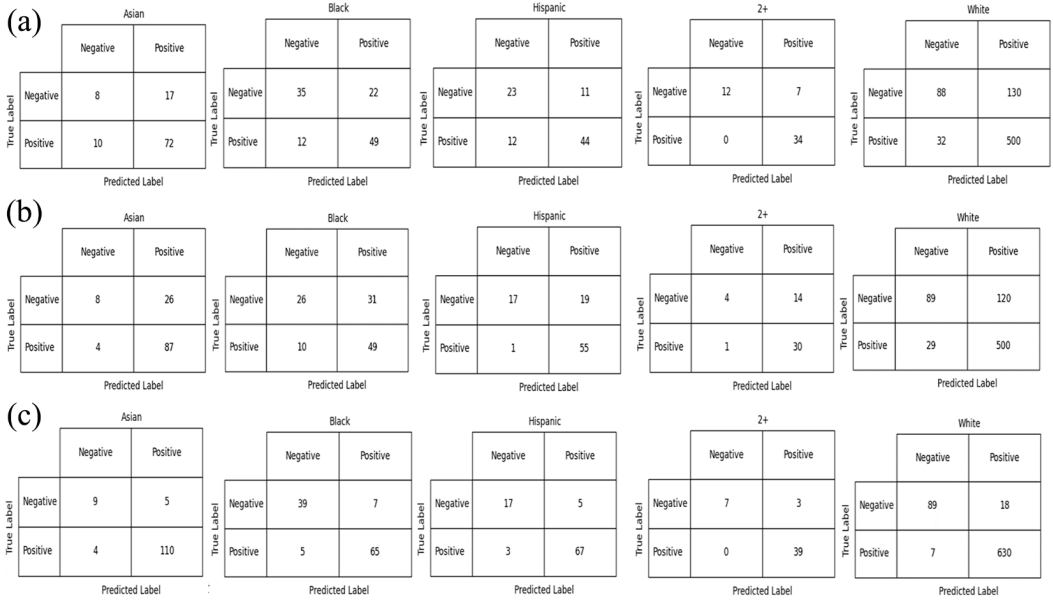

FIGURE 2. *Confusion matrices for all racial/ethnic groups. (a) RF confusion matrices of baseline, (b) RF confusion matrices of ExGR, and (c) RF confusion matrices of MetaC.*

The confusion matrices in Figure 2a present concrete examples of model accuracy. With respect to statistical parity, students in the Asian and White categories have, respectively, 89% and 84% probability of attainment, while those in the Black and Hispanic categories have 60% and 61% probabilities. Without correcting for bias, predictive models will be more likely to predict that students categorized as Black and Hispanic are less likely to attain a bachelor's degree or higher than their more privileged peers.

The findings for predictive equality illustrate bias in the *accuracy* of the predictions. Among the students who did not complete their degree ($y = 0$), the probability of completing their degree (predicted success, despite actual failure) is estimated as 65% for White and 73% for Asian, while it is estimated as 28% for Hispanic and 33% for Black (Figure 2a). As illustrated in Figure 1b, the models are also more likely to falsely predict failure for Black and Hispanic students than for White and Asian students. Illustratively, among the students who completed their degree ($y = 1$), the probability of failure is estimated as 12% for White and 6% for Asian students, while it is estimated as 21% for Hispanic and 19% for Black students, as illustrated in Figure 2a.

Moreover, the boxplots show that the variation of values for the White and Asian groups is minimal, especially for the White group, whereas the variation of unfairness gaps for the other groups is significantly larger. Differences in variation between racial/ethnic groups indicate that models for minoritized groups are more sensitive to train/test splits. Due to the population bias across different racial/ethnic groups in the ELS dataset (i.e., statistical underrepresentation of Black and Hispanic students), the train/test splits can significantly change the distribution and presence of underrepresented individuals in each partition, significantly impacting the unfairness of the model for each split scenario. In practice, this will result in less stable and fair model performance for predicting the success of an unobserved individual from a statistically underrepresented group.

*Aggregate Level: Privileged vs. Unprivileged.* Figure 3 presents the box plots for all four unfairness notions at the aggregated level of privileged (Asian and White) versus unprivileged (Black, Hispanic, and two or more racial/ethnic categories) for all prediction models. The first evident pattern from all four plots is the mean difference between the two groups across all four fairness notions. Similar to results at the subgroup level, we observe worse model performance (accuracy) for the unprivileged group. This accuracy disparity manifests as higher false negative rates for the unprivileged group and lower false positive rates for the unprivileged group.

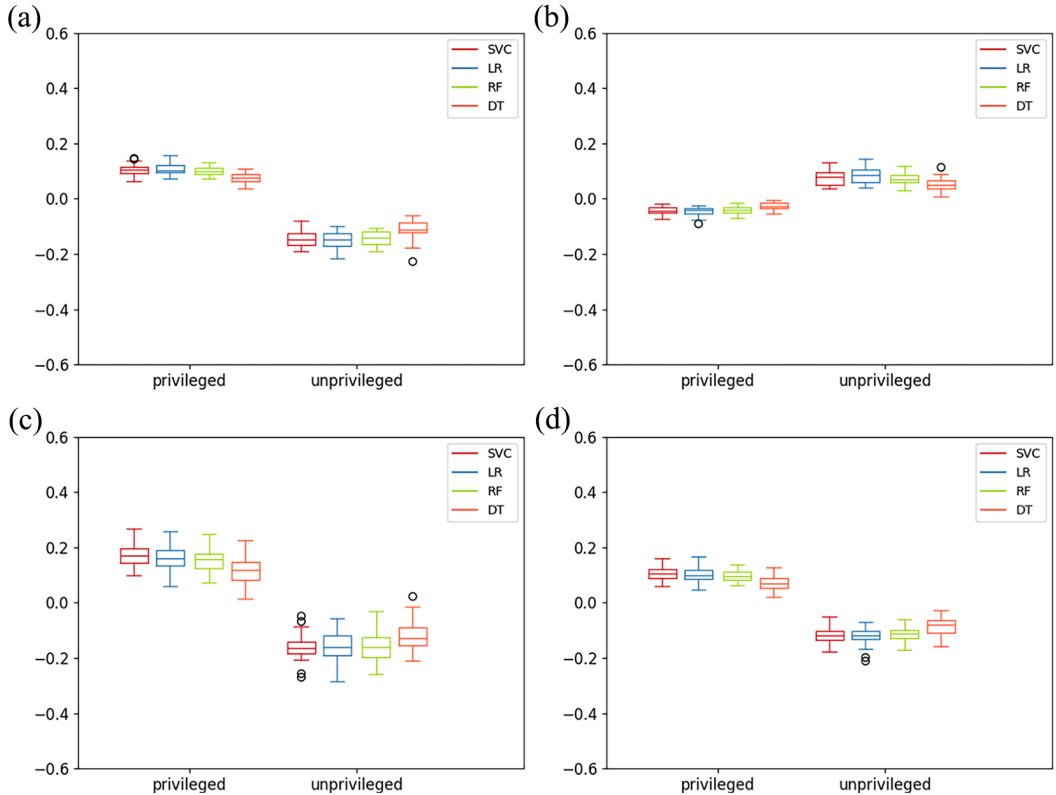

FIGURE 3.  *Baseline with all ML models for privileged vs. unprivileged groups. (a) Statistical parity of baseline for different ML models, (b) equal opportunity of baseline for different ML mod, (c) predictive equality of baseline for different ML models, and (d) equalized odds of baseline for different ML models.*

Comparing findings at the subgroup and aggregate levels, we observe that aggregate results mask substantial differences we can glean from the subgroup analysis. For instance, in Figure 1a, all the models demonstrate greater unfairness toward the Black group rather than the Hispanic group, especially in the case of SVM and LR. At the aggregate level of analysis, however, this variation cannot be observed (all models are unfair toward the unprivileged group). Next, we turn to the results for bias-mitigation techniques.

*Mitigating Bias*

Given space constraints and for ease of interpretability, we present mitigation results using one predictive model, RF, which is a nonlinear classifier commonly used in the education literature. These results appear in Figure 4 (findings from other ML models are in Appendix F). Our first observation is that the preprocessing and in-processing mitigation methods only minimally decrease accuracy (by 1% to 2%). One technique, MetaC, significantly improves accuracy (by 10 to 17 points over the baseline model without bias mitigation).

The bias-mitigation techniques we used required us to specify the privileged and unprivileged groups and to treat the sensitive attribute as binary. The results indicate that certain techniques demonstrate greater effectiveness than others, particularly when paired with specific ML models or fairness measures. We first present findings for the preprocessing mitigation techniques, ReW and DIR.

The results (in Figure 4) indicate that the ReW technique is minimally effective in mitigating bias compared to the baseline case. If the goal of the education data analyst is to reduce unfairness in student success predictions, increasing the influence of data points that represent successful students from unprivileged groups (e.g., Black students who succeed) in the training process is not enough. This finding suggests that the underrepresentation of successful students from unprivileged groups in the training data is not a key source of bias in student success predictions.

Upon further analysis (in Appendix F), it is evident that ReW performs better with linear ML models like SVM or LR compared to nonlinear models such as RF or DT. This is likely because ReW, where different weights are assigned to observations, adjusts the data distribution, enhancing linear separability and allowing linear models to leverage the refined data distribution effectively.

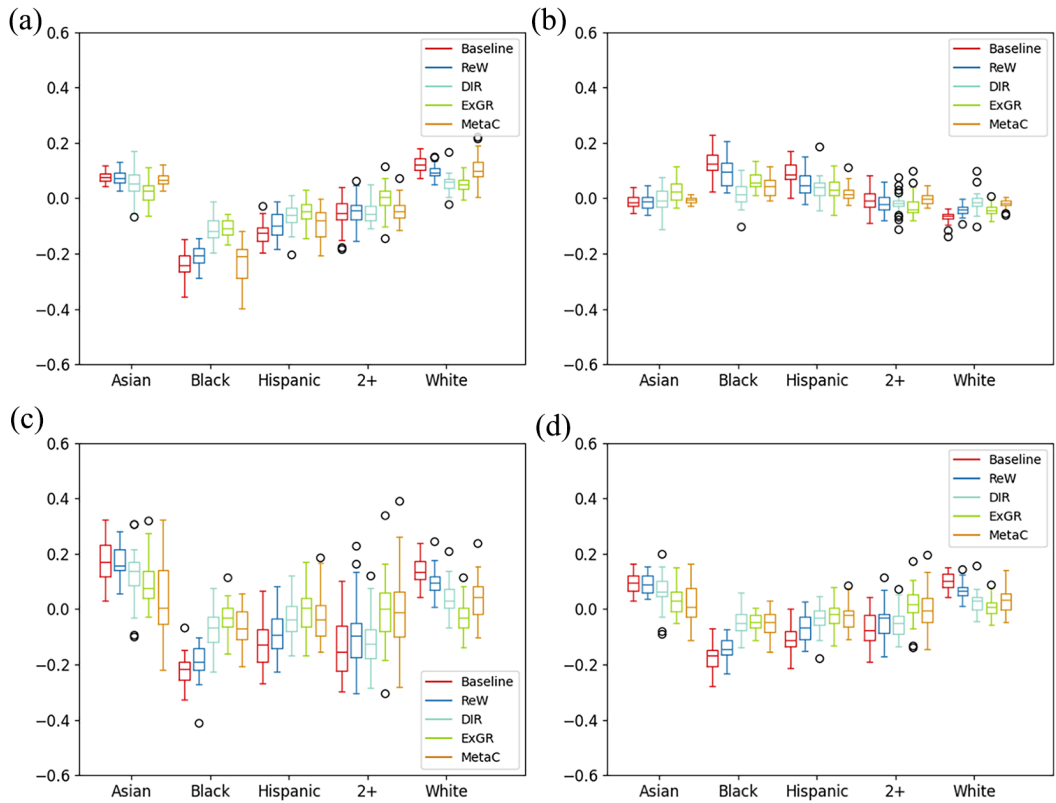

FIGURE 4. *Mitigation for all racial/ethnic groups. (a) Statistical parity of RF using bias mitigation techniques, (b) equal opportunity of RF using bias mitigation techniques, (c) predictive equality of RF using bias mitigation techniques, and (d) equalized odds of RF using bias-mitigation techniques.*

The second preprocessing mitigation technique we employed, DIR, effectively reduces model unfairness for all the unprivileged groups, demonstrating notable effectiveness for the Black group. This approach modifies the distributions of other features in the model (e.g., students' native language and family composition) to reduce their correlation with racial/ethnic categorizations. A feature can provide a strong indication of the group to which a data point might belong. DIR aims to eliminate this capacity to distinguish between group membership. In addition to reducing unfairness for the Black group, DIR diminishes the modeling bias in favor of the White group relative to other groups. However, the bias in favor of the Asian group is only partially diminished, and it does not substantially improve fairness across all fairness notions for this specific subgroup. In line with expectations, applying DIR decreases the equal opportunity gap between the White group and all other groups, indicating a decrease in the number of successful students from unprivileged groups who are falsely predicted to be unsuccessful.

Note that the DIR approach corrects the dataset measuring and considering the statistical parity notion at the aggregate level. Hence, it is expected to observe equal proportions of positive prediction from each group at the aggregated level of privileged versus unprivileged. However, our results show that DIR does not fully achieve statistical parity for each subgroup using ELS data. At the aggregate level (Figure 5), DIR slightly removes the modeling bias in favor of the privileged group and reduces the bias against the unprivileged group, although not as effectively as other mitigation techniques. Findings for DIR at the subgroup level (Figure 4) also highlight differences between two groups that are often considered privileged (Asian and White) and two groups that are often considered unprivileged (Black and Hispanic), underscoring the importance of disaggregation.

Contrary to ReW findings, results in Appendix F demonstrate that DIR performs better with nonlinear models like RF or DT rather than linear models. This outcome may stem from the adjustments made by DIR, which aim to reduce disparate impact but may lead to a dataset requiring complex decision boundaries for effective class separation. Nonlinear models are better at dealing with complexities because they can more effectively capture intricate relationships in the data.

Turning to the in-processing techniques, ExGR emerges as the most effective technique in reducing bias. Although it does not fully eliminate disparities, it significantly mitigates the model bias in favor of the White group and reduces the bias against the Black and Hispanic groups across all four fairness notions. This is likely because the ExGR, as an in-processing method, integrates fairness constraints directly into the training process, using a specific fairness metric as a constraint incorporated into the optimization objective function during the execution of ExGR.

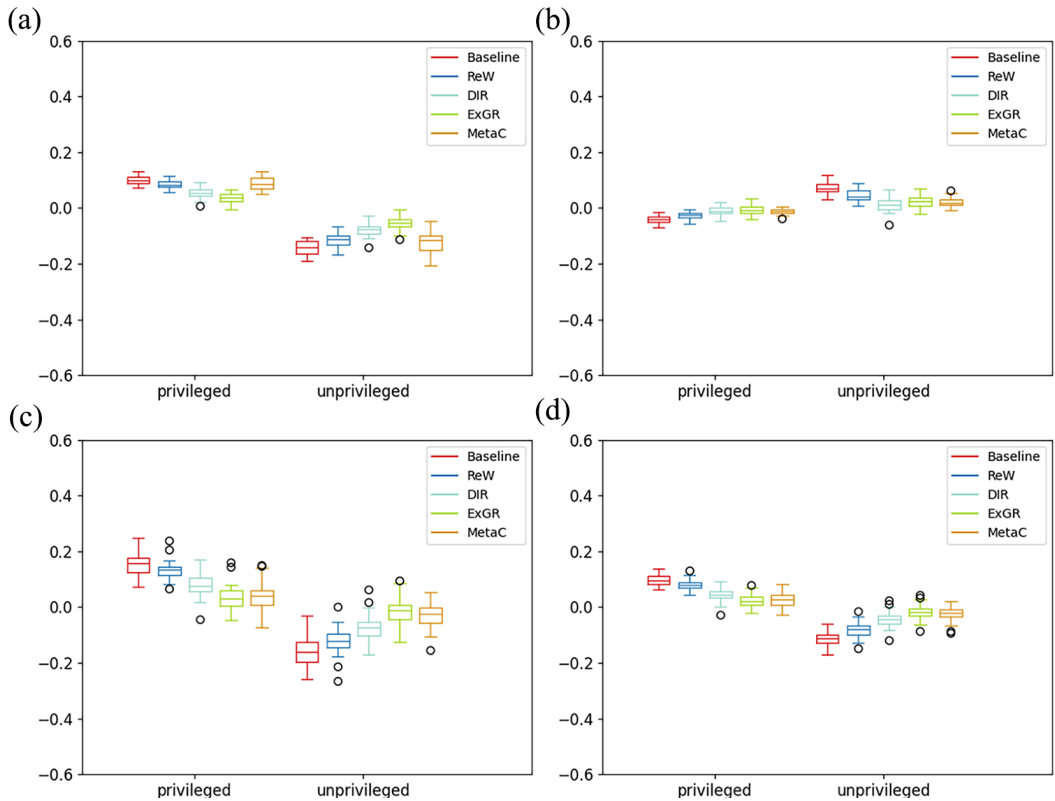

FIGURE 5. *Mitigation for privileged vs. unprivileged groups. (a) statistical parity of RF using bias mitigation techniques, (b) equal opportunity of RF using bias mitigation techniques, (c) predictive equality of RF using bias mitigation techniques, and (d) equalized odds of RF using bias-mitigation techniques.*

The MetaC technique effectively mitigates biases for both the Hispanic and Black groups concerning equal opportunity, predictive equality, and equalized odds, the fairness notions related to prediction accuracy. However, MetaC does not enhance fairness in the case of statistical parity. Moreover, MetaC leads to increased variability for predictive equality and equalized odds, suggesting that the adjusted model is less robust to variations in data splits.

Results at the subgroup level can be further explored by examining the confusion matrices in Figure 2 (findings from other techniques are in Appendix G). These results were computed by aggregating across 30 different splits for each race and mitigation technique combination. Notably, ExGR demonstrates superior performance with Hispanic students compared to Black students. Particularly, it shows a greater reduction compared to the baseline in the number of successful Hispanic students falsely predicted to be unsuccessful. MetaC, in alignment with its higher accuracy, enhances the number of correctly predicted students for both the Black and Hispanic groups. Additionally, it is noteworthy that it significantly decreases the number of unsuccessful White students falsely predicted to be successful, reducing disparities in predictions between White students and other groups.

The results confirm that even at the aggregated level, unfairness is only partially mitigated, with ExGR emerging as the most effective technique. The results suggest that in-processing techniques, while not achieving perfect performance, are more effective in reducing bias compared to preprocessing techniques. Future work should examine bias mitigation when both preprocessing and in-processing techniques are applied simultaneously.

**Discussion**

This study sought to examine how college student success predictions, including the accuracy of predictions, differ across racial/ethnic groups. We also evaluated the effectiveness of four common bias-mitigation techniques. Using a nationally representative dataset with student-level data, we demonstrate how prediction models yield less accurate results for Black and Hispanic students, systematically making more errors in predictions for these students. These models are significantly more likely to predict failure for Black and Hispanic students who actually succeeded and less likely to predict success for students who failed compared to White and Asian groups.

In addition, we illustrate both the potential and the limitations of existing techniques in mitigating bias. Bias-mitigation techniques generally show promise in improving fairness, but all fail to fully eliminate disparities in the predictions across racial/ethnic groups. In-processing techniques, those that account for fairness in the modeling process, are generally more effective at mitigating algorithmic bias across different ML models compared to preprocessing techniques, which adjust the data before modeling.

The practical implications of these findings are significant but depend on the use of prediction outcomes. For example, if such models are used to make college admissions decisions, admission may be denied to racially minoritized students if the models show that previous students with the same racial categorization have lower predicted likelihoods of success. Considering mounting evidence of algorithmic bias, this potential consequence has led researchers to caution that such models should not be used for admission decisions (Hutt et al., 2019).

Likewise, with course recommendations, higher education observers have warned that predictions could lead to educational tracking, encouraging students from racially minoritized groups to pursue courses or majors that are perceived as less challenging (Ekowo & Palmer, 2017). Such consequences may go undetected since automated sorting mechanisms remain both obfuscated (due to their invisibility to educational stakeholders) and legitimized through perceptions that statistical models are objective (Hirschman & Bosk, 2020).

In contrast to the previous scenarios, biased models may lead to greater support for disadvantaged students when used to inform student success interventions. By falsely predicting failure for racially minoritized students who succeed, the model may direct greater resources to those students the model is biased against. Even then, practitioners must be careful not to produce deficit narratives of minoritized students, treating them as though they have a lower probability of success.

The evidence of algorithmic bias, coupled with varying implications depending on the use of predictions, signals the importance of training end users on the potential for algorithmic bias. End users comprise a wide array of college and university staff members, including admissions officials, academic advisors, and faculty members (Algarni & Sheldon, 2023; Chen et al., 2020; Denley, 2013; Klempin et al., 2018; Tough, 2021). Awareness of potential algorithmic bias, including its direction and the groups targeted by the bias, can help users contextualize predictions for individual students and make more informed decisions.

Beyond training end users, the findings from this study point to the potential value of employing statistical techniques to mitigate algorithmic bias and leveling model performance across groups. While the techniques we employed failed to eliminate disparities in prediction outcomes or accuracy, some techniques effectively reduced bias in the models, particularly those that incorporate fairness constraints in the modeling process. Analysts building predictive models, either within institutions or for third-party vendors, should explore the use of bias-mitigation techniques, particularly given growing concerns over algorithmic bias (Barocas & Selbst, 2016; Ekowo & Palmer, 2017; Kizilcec & Lee, 2022).

As higher education institutions strive to better serve students by becoming more data-informed (Gagliardi & Turk, 2017), it is imperative that predictive models are designed with attention to their potential social consequences. It is critical to be aware of historical discrimination reflected in the data and consider fairness measures to reduce bias in the outcomes of the models. This paper demonstrates that more work is needed to reduce algorithmic bias across racialized groups. Future research should examine the effectiveness of other bias-mitigation techniques. Another important avenue for future work is understanding how feature selection (which variables to include in the model) affects prediction accuracy and fairness across racialized groups. Such work could expand on existing and conflicting recommendations concerning the inclusion of race/ethnicity variables in student success prediction models (Baker et al., 2023; Hu & Rangwala, 2020; Yu et al., 2021). Finally, while we demonstrate the importance of disaggregating beyond privileged/unprivileged, the ELS racial/ethnic categories are severely limited. Future work should disaggregate further to lead us toward more racially just student-success practices in higher education.

## Declaration of Conflicting Interests

The author(s) declared no potential conflicts of interest with respect to the research, authorship, and/or publication of this article.

## Funding

The author(s) disclosed receipt of the following financial support for the research, authorship, and/or publication of this article: The research reported here was supported, in whole or in part, by the Institute of Education Sciences, U.S. Department of Education, through grant R305D220055 to the University of Illinois Chicago and by grant P2CHD042849, awarded to the Population Research Center at The University of Texas at Austin by the Eunice Kennedy Shriver National Institute of Child Health and Human Development. The content is solely the responsibility of the authors.

## ORCID iDs

Denisa Gándara  https://orcid.org/0000-0001-5714-5583

Hadis Anahideh  https://orcid.org/0000-0003-1935-7571

## Notes

1. In ML, a training dataset includes the data you use to train the model or algorithm to predict the outcome of interest.

2. An examination of the impact of including specific features in predictive models is beyond the scope of this study. Such work, under progress, includes the review of a rich literature and discussion of other important considerations, including social and political contexts. Such background and analysis cannot receive adequate treatment in conjunction with the analyses presented in this study.

3. In all versions, we avoid imputing socially relevant (sensitive) attributes and outcome variables. Hence, observations with missing values for these variables are always dropped before imputation.

4. The F1-score is employed to compute metrics for individual labels and obtain their weighted average based on support, which accounts for label imbalance by considering the number of true instances for each label.

**Authors**

DENISA GÁNDARA is an assistant professor at The University of Texas at Austin, George I. Sánchez Bldg, 1912 Speedway, Austin, TX 78705; denisa.gandara@austin.utexas.edu. Her research focuses on higher education finance, policy, and politics.

HADIS ANAHIDEH is an assistant professor at the University of Illinois Chicago, 842 W. Taylor Street, Chicago, Illinois 60607; email: hanahide@uic.edu. Her research focuses on developing innovative data-driven learning and optimization algorithms, with a special emphasis on the ethical implications of these technologies.

MATTHEW ISON (he/him) is an instructor in the counseling and higher education program at Northern Illinois University, Barsema Alumni and Visitors Center, 231 N Annie Glidden Rd., Dekalb, IL 60115; mison@niu.edu. His research focuses on the role of federal, state, and institutional policy in shaping access and retention for community college students.

LORENZO PICCHIARINI is a student at the University of Illinois Chicago, 842 W. Taylor Street, Chicago, IL 60607; email: lpicch3@uic.edu. His research focuses on developing fairness-aware modeling for classification tasks to reduce disparities toward specific individuals or groups.